\documentstyle[12pt]{article}

\begin{document}

\raggedbottom
\baselineskip=15pt
\parskip=15pt
\pagestyle{empty}
\begin{center}
{\Large\bf Inhomogeneous and anisotropic cosmologies}
\vspace*{5mm} \\
{\large M.A.H. MacCallum}
\vspace*{15pt} \\
School of Mathematical Sciences \\
Queen Mary and Westfield College, University of London \\
Mile End Road, London E1 4NS, U.K. \\
E-mail: M.A.H.MacCallum@qmw.ac.uk
\end{center}

\section{Introduction}

Although the purpose of this workshop is to discuss the structures in
the universe, which are inhomogeneous, homogeneous models have been
used in considering many of the cosmological issues raised in that
discussion, so I have also included in this survey the anisotropic
homogeneous models and their implications.  Only exact solutions will
be covered: other speakers at Pont d'Oye (e.g.\ Bardeen,
Brandenberger, Dunsby and Ellis) gave very full discussions of
perturbation theory.  Here I continue my previous practice (MacCallum
1979, 1984)\nocite{Mac79,Mac84}, by using the mathematical
classification of the solutions as an overall scheme of organization;
the earlier reviews give additional details and references. (A survey
organized by the nature of the applications is to appear in the
proceedings of Dennis Sciama's 65th birthday meeting.) Other useful
reviews are: Ryan and Shepley (1975)\nocite{RyaShe75} on homogeneous
anisotropic models; Krasinski (1990)\nocite{Kra90} on inhomogeneous
models; and Verdaguer (1985, 1992)\nocite{Ver85,Ver92} on models of
solitonic character.

In section 2 I will consider the spatially-homogeneous but anisotropic
models. These are the Bianchi models, in general, the exceptions being
the Kantowski-Sachs models with an $S^2 \times R^2$ topology. Such models
could be significant in understanding the background in which
structure is formed, but they do not themselves model that structure.
However, I will include here some remarks about inhomogeneous models
which are closely related to calculations done with Bianchi models.
Then in section 3 I will consider the inhomogeneous models, which fall
into several classes. They can be used both as local models of
structure and as possible global models of the background in which
structure forms (and are in some cases used for both purposes
simultaneously). A final section attempts a synthesis and makes some
summarizing remarks.
\vspace*{15pt}

\parskip=0pt
 What is it that a cosmological model should explain? There are
the following main features:
 \begin{list}{[\arabic{enumi}]}{\setlength{\itemsep}{0pt}
  \setlength{\partopsep}{0pt}  \setlength{\topsep}{0pt}
  \setlength{\parsep}{0pt}\usecounter{enumi}}
 \item{Lumpiness, or the clumping of matter. The evidence for this is
obvious.}
 \item{Expansion, shown by the Hubble law.}
 \item{Evolution, shown by the radio source counts and more recently
by galaxy counts.}
 \item{A hot dense phase, to account for the cosmic microwave
background radiation (CMWBR) and the abundances of the chemical elements.}
 \item{Isotropy, shown to a high degree of approximation in various
cosmological observations, but especially in the CMWBR.}
 \item{Possibly, homogeneity. (The doubt indicated here will be
explained later.)}
 \item{The numerical values of parameters of the universe and its
laws, such as the baryon number density, the total density parameter
$\Omega$, the entropy per baryon, and the coupling constants}
 \item{(Perhaps) such features as the presence of life.}
 \end{list}
\vspace*{15pt}

Originally, the standard big-bang models were the
Friedman-Lema\^{\i}tre-Robertson-Walker (FLRW) models characterized as:
 \begin{list}{[\arabic{enumi}]}{\setlength{\itemsep}{0pt}
  \setlength{\partopsep}{0pt}  \setlength{\topsep}{0pt}
  \setlength{\parsep}{0pt}\usecounter{enumi}}
 \item{Isotropic at all points and thus necessarily\ldots}
 \item{Spatially-homogeneous, implying Robertson-Walker geometry.}
 \item{Satisfying Einstein's field equations}
 \item{At recent times (for about the last $10^{10}$ years)
pressureless and thus governed by the Friedman-Lema\^{\i}tre dynamics.}
 \item{At early times, radiation-dominated, giving the Tolman dynamics
and a thermal history including the usual account of nucleogenesis and
the microwave background.}
 \end{list}
\vspace*{15pt}

To this picture, which was the orthodox view from about 1965-80, the
last decade has added the following extra orthodoxies:
 \begin{list}{[\arabic{enumi}]}{\setlength{\itemsep}{0pt}
  \setlength{\partopsep}{0pt}  \setlength{\topsep}{0pt}
  \setlength{\parsep}{0pt}\usecounter{enumi}}
 \setcounter{enumi}{5}
 \item{$\Omega = 1$. Thus there is dark matter, for which the Cold
Dark Matter model was preferred.}
 \item{Inflation -- a period in the early universe where some field
effectively mimics a large cosmological constant and so causes a
period of rapid expansion long enough to multiply the initial length
scale many times.}
 \item{Non-linear clustering on galaxy cluster scales, modelled by the
$N$-body simulations which fit correlation functions based on observations.}
 \end{list}
and also added, as alternatives, such concepts as cosmic strings, GUTs or
TOEs\footnote{Why so anatomical?} and so on.
\parskip=15pt

The standard model has some clear successes: it certainly fits the
Hubble law, the source count evolutions (in principle if not in
detail), the cosmic microwave spectrum, the chemical abundances, the
measured isotropies, and the assumption of homogeneity. Perhaps its
greatest success was the prediction that the number of neutrino
species should be 3 and could not be more than 4, a prediction now
fully borne out by the LEP data.

However, the model still has weaknesses \cite{Mac87}.  For example,
the true clumping of matter on large scales, as shown by the QDOT data
\cite{SauFreRow91} and the angular correlation functions of galaxies
\cite{MadEfsSut90}, is too strong for the standard cold dark matter
account\footnote{These discoveries made it possible for disagreement
with the 1980s dogmatism on such matters to at last be listened to.}.
The uniformity of the Hubble flow is under question from the work of
the ``Seven Samurai'' \cite{LynFabBur88} and others. The question of
the true value of $\Omega$ has been re-opened, partly because theory
has shown that inflation does not uniquely predict $\Omega\:=\:1$
(cf.\ Ellis' talk at Pont d'Oye) and partly because observations give
somewhat variant values. Some authors have pointed out that our
knowledge of the physics valid at nucleogenesis and before is still
somewhat uncertain, and that we should thus retain some agnosticism
towards our account of those early times.

 Finally, we should recognize that our belief in homogeneity on a
large scale has very poor observational support. We have data from our
past light cone (and those of earlier human astronomers) and from
geological records \cite{Hoy62}.  Studying spatial homogeneity requires us to
know about conditions at great distances {\em at the present time},
whereas what we can observe at great distances is what happened a long
time ago, so to test homogeneity we have to understand the evolution
both of the universe's geometry and of its matter content%
\footnote{Local measures of homogeneity merely tell us that the
spatial gradients of cosmic quantities are not too strong near us.}.
Thus we cannot test homogeneity, only check that it is consistent with
the data and our understanding of the theory.  The general belief in
homogeneity is indeed like the zeal of the convert, since until the
1950s, when Baade revised the distance scale, the accepted distances
and sizes of galaxies were not consistent with homogeneity.

 These comments, however, are not enough to justify examination of
other models. Why do we do that? The basic reason is to study
situations where the FLRW models, even with linearized perturbations,
may not be adequate. Three types of situation come to mind: the fully
non-linear modelling of local processes; exploration of the uniqueness
of features of the FLRW models; and tests of the viability of non-FLRW
models. The uniqueness referred to here may lie in characteristics
thought to be peculiar to the standard model; in attempted proofs that
no model universe could be anisotropic or inhomogeneous, by proving
that any strong departures from the standard model decay away during
evolution; or in comparisons with observation, to show that only the
standard models fit.

 Some defects of the present survey should be noted.  One is that the
matter content is generally assumed to be a perfect fluid, although
this is strictly incompatible with the other assumed physical
properties. Attempting to remedy this with some other mathematically
convenient equation of state is not an adequate response; one must try
to base the description of matter on a realistic model of microscopic
physics or thermodynamics, and few have considered such questions
\cite{BraSvi84,SalSchLee87,BonCol88,RomPav92}.

A second limitation is that we can only explore the mathematically
tractable subsets of models\footnote{Kramer {\em et al.\/} (1980)
provides a detailed survey of those classes of relativistic spacetimes
where the Einstein field equations are sufficiently tractable to be
exactly solved.}, which may be far from representative of all models.
To avoid this restriction, we will ultimately have to turn to
numerical simulations, including fully three-dimensional variations in
the initial data. Some excellent pioneering work has of course been
done, e.g.\ Anninos {\em et al.\/} (1991b)\nocite{AnnMatTul91}, but
capabilities are still limited (for example Matzner (1991)\nocite{Mat91}
could only use a space grid of $31^3$ points and 256
time steps). Moreover, before one can rely on numerical simulations
one needs to prove some structural stability results to guarantee that
the numerical and exact answers will correspond.

As a final limitation, in giving this review I only had time to
mention and discuss some selected papers and issues, not survey the
whole vast field. For his mammoth survey of all inhomogeneous
cosmological models which contain, as a limiting case, the FLRW
models, Krasinski now has read about 1900 papers (as reported at the
GR13 conference in 1992)\footnote{The survey is not yet complete and
remains to be published, but interim reports have appeared in some
places, e.g.\ Krasinski (1990)\nocite{Kra90}.}.  Thus the bibliography
is at best a representative selection from many worthy and interesting
papers, and authors whose work is unkindly omitted may quite
reasonably feel it is {\em un}representative. In particular, I have
not attempted to cover the higher-dimensional models discussed by
Demaret and others.

\section{Spatially-homogeneous anisotropic models}

\subsection{Metrics and field equations}

As already mentioned, this class consists of the Bianchi and
Kantowski-Sachs models.  They have the advantage that the Einstein
equations reduce to a system of ordinary differential equations,
enabling the use of techniques from dynamical systems theory, and
there is thus again a vast literature, too big to fully survey here.

The Bianchi models can be defined as spacetimes with metrics
 $$
 ds^2 = -dt^2 + g_{\alpha\beta}(t)({e^\alpha}_\mu dx^\mu)({e^\beta}_\nu dx^\nu)
 $$
where the corresponding basis vectors $\{{\bf e}_\alpha\}$ obey
 $$
 [{\bf e}_\alpha,\:{\bf e}_\beta ] = {C^\gamma}_{\alpha\beta}{\bf e}_\gamma
 $$
 in which the C's are the structure constants of the relevant symmetry
group. The different Bianchi-Behr types I-IX are then defined (see e.g.\
Kramer {\em et al.\/} (1980)) by
algebraic classification of these sets of structure constants.

The Kantowski-Sachs metric is
 $$
 ds^2 = -dt^2 + a^2(t)\,dx^2 + b^2(t)(d\theta^2 +\sin^2 \theta \, d\phi^2).
 $$
(The other metric given in the original paper of Kantowski and Sachs
was in fact a Bianchi metric, as pointed out by Ellis.)

The adoption of methods from the theory of dynamical systems has
considerably advanced the studies of the behaviour of Bianchi models,
beginning in the early 70s with the discussion of phase portraits for
special cases \cite{Col71}. Subsequently, more general cases were
discussed using a compactified phase space.
In the last decade these methods have been coupled with the
parametrization of the Bianchi models using automorphism group
variables \cite{ColHaw73,Har79,Jan79,Sik80,RoqEll85,Jak87}.

The automorphism group can be briefly described as follows. Take
a transformation
 $$ \hat{{\bf e}}^\alpha = {M^\alpha}_\beta {\bf e}^\beta. $$
 This is an automorphism of the symmetry group if the
 $\{\hat{{\bf e}}_\alpha\}$ obey the same commutation relations as the
$\{{\bf e}_\alpha\}$. The matrices $M$ are time-dependent and are
chosen so that the new metric coefficients $\hat{g}_{\alpha\beta}$
take some convenient form, for example, become diagonal. The real
dynamics is in these metric coefficients. This idea is present in
earlier treatments which grew from Misner's methods for the Mixmaster
case \cite{RyaShe75} but unfortunately the type IX case was highly
misleading in that for Bianchi IX (and no others except Bianchi I) the
rotation group is an automorphism group.

The compactification of phase space, introduced for general cases by
S.P. Novikov and Bogoyavlenskii (see Bogoyavlenskii (1985))
entailed the normalization of configuration variables to lie within some
bounded region, which was then exploited by (a) finding Lyapunov
functions, driving the system near the boundaries of the phase space and (b)
using analyticity, together with the behaviour of critical points and
separatrices, to derive the asymptotic behaviour.

Three main groups have developed these treatments: Bogoyavlenskii and
his colleagues ({\em op. cit}); Jantzen, Rosquist and collaborators
(e.g.\ Jantzen (1984), Rosquist {\em et al.\/} (1990)%
\nocite{Jan84,RosUggJan90}) who have coupled the automorphism
variables with Hamiltonian treatments in a powerful formalism; and
Wainwright and colleagues (e.g.\ Wainwright and Hsu (1989)%
\nocite{WaiHsu89}) who have used a different, and in some respects
simpler, set of automorphism variables, which are well-suited for
studying asymptotic behaviour because their limiting cases are
physical evolutions of simpler models rather than singular behaviours.
Similar ideas can be used for the Kantowski-Sachs models too. As well
as qualitative results, some of them described below, these methods
have enabled new exact solutions to be found, and some general
statements about the occurrence of these solutions to be made.

Many of the geometrical properties of Bianchi cosmologies can be
carried over to cases where the 3-dimensional symmetry group (which is
still classifiable by Bianchi type) acts on timelike surfaces. A
number of authors have considered such metrics, for example Harness
(1982)\nocite{Har82} and myself and Siklos (1992)\nocite{MacSik92}.
Although of less interest, since they do not evolve in time, than the
usual Bianchi models, some of these models reappear as (spatially)
inhomogeneous static or stationary models below.

Since the present-day universe is not so anisotropic that we can
readily detect its shear and vorticity, the Bianchi models can be
relevant to cosmology only as models of asymptotic behaviour, in the
early or late universe, or over long time scales, such as the time
since the ``last scattering''. They have also been used, in these
contexts, as approximations in genuinely inhomogeneous universes, and
one has to be careful to distinguish the approximate and exact uses.

\subsection{Asymptotic behaviour: the far past and future}

The earliest use of anisotropic cosmological models to study a real
cosmological problem was the investigation by Lema\^{\i}tre (1933)%
\nocite{Lem33} of the occurrence of singularities in Bianchi type I
models. The objective was to explore whether the big-bang which arose
in FLRW models was simply a consequence of the assumed symmetry: it
was of course found not to be.

One can argue that classical cosmologies are irrelevant before the
Planck time, but until a theory of quantum gravity is established and
experimentally verified (if indeed that will ever be possible) there
will be room for discussions of the behaviour of classical models near
their singularities.

In the late 1950s and early 60s Lifshitz and Khalatnikov and their
collaborators showed (a) that singularities in synchronous coordinates
in inhomogeneous cosmologies were in general `fictitious' and (b) that
a special subclass gave real curvature singularities, with an
asymptotic behaviour like that of the Kasner (vacuum Bianchi I)
cosmology \cite{LifKha63}.  From these facts they (wrongly) inferred
that general solutions did not have singularities. This contradicted
the later singularity theorems (for which see Hawking and Ellis
(1973)\nocite{HawEll73}), a disagreement which led to the belief that
there were errors in LK's arguments.  They themselves, in
collaboration with Belinskii, and independently Misner, showed that
Bianchi IX models gave a more complicated, oscillatory, behaviour than
had been discussed in the earlier work, and Misner christened this the
`Mixmaster' universe after a brand of food mixer. The broad picture of
the r\^{o}les of the Kasner-like and oscillatory behaviours has been
borne out by the more rigorous studies by the methods described in the
previous section. There is also an interesting and as yet incompletely
explored result that after the oscillatory phase many models
approximate one of a few particular power-law (self-similar) solutions
\cite{Bog85}.

The detailed behaviour of the Mixmaster model has been the subject of
still-continuing investigations: some authors argue that the evolution
shows ergodic and chaotic properties, while others have pointed out
that the conclusions depend crucially on the choice of time variable
\cite{Bar82,BurBurEll90,Ber91}. Numerical investigations are
tricky because of the required dynamic range if one is to study an
adequately large time-interval, and the difficulties of integrating
chaotic systems.

The extension of these ideas to the inhomogeneous case, by Belinskii,
Lifshitz and Khalatnikov, has been even more controversial, though
prompting a smaller literature. It was strongly attacked by Barrow and
Tipler (1979)\nocite{BarTip79} on a number of technical grounds, but
one can take the view that these were not as damaging to the case as
Barrow and Tipler suggested \cite{BelLifKha80,Mac82}.  Indeed the
`velocity-dominated' class whose singularities are like the Kasner
cosmology have been more rigorously characterized and the results
justified \cite{EarLiaSac71,HolJolSma90}.  Sadly this does not settle
the more general question, and attempts to handle the whole argument
on a completely rigorous footing\footnote{One of them made by
Smallwood and myself.} have so far failed.

General results about singularity types have been proved. The `locally
extendible' singularities, in which the region around any geodesic
encountering the singularity can be extended beyond the singular
point, can only exist under strong restrictions \cite{Cla76},
while the `whimper' singularities \cite{KinEll73}, in which
curvature invariants remain bounded while curvature components in some
frames blow up, have been shown to be non-generic and unstable
\cite{Sik78}. Examples of these special cases were found among
Bianchi models, and both homogeneous and inhomogeneous cosmologies
have been used as examples or counter-examples in the debate.

A further stimulus to the study of singularities was provided by
Penrose's conjecture that gravitational entropy should be low at the
start of the universe and this would correspond to a state of small or
zero Weyl tensor \cite{Pen79,Tod92}.

Many authors have also considered the far future evolution (or, in
closed models, the question of recollapse, whose necessity in Bianchi
IX models lacked a rigorous proof until recently \cite{LinWal91}).
{}From various works \cite{Mac71,ColHaw73,BarTip78} one finds that the
homogeneous but anisotropic models do not in general settle down to an
FLRW-like behaviour but typically generate shears of the order of 25\%
of their expansion rates; see also \cite{UggJanRos91}.  From the
dynamical systems treatments, it is found that certain exact solutions
(which in general have self-similarity in time) act as attractors of
the dynamical systems in the future \cite{WaiHsu89}.  (All such exact
solutions are known: see Hsu and Wainwright (1986)\nocite{HsuWai86}
and Jantzen and Rosquist (1986)\nocite{JanRos86}.)

This last touches on an interesting question about our account of the
evolution of the universe: is it structurally stable, or would small
changes in the theory of the model parameters change the behaviour
grossly? Several instances of the latter phenomenon, `fragility', have
recently been explored by Tavakol, in collaboration with Coley, Ellis,
Farina, Van den Bergh and others \cite{ColTav92}.

\subsection{Long time effects: the cosmic microwave background}

To test the significance of the observed isotropy of the CMWBR, many
people in the 1960s and 70s computed the angular distribution of the
CMWBR temperature in Bianchi models (e.g.\ Thorne (1967), Novikov
(1968), Collins and Hawking (1972), and Barrow {\em et al.\/}
(1983)\nocite{Tho67,Nov68,ColHaw72,BarJusSon83}). These calculations
allow limits to be put on small deviations from isotropy from
observation, and also enabled, for example, the prediction of `hot
spots' in the CMWBR in certain Bianchi models, which could in
principle be searched for, if there were a quadrupole component, as
there is in the COBE data (though perhaps not for this reason), to see
if the quadrupole verifies one of those models.

Similar calculations, by fewer people, considered the polarization
\cite{Ree68,Ani74,TolMat84} and spectrum \cite{Ree68,Ras71}. More
recently still, work has been carried out on the microwave background
in some inhomogeneous models \cite{SaeArn90}. It has been shown that pure
rotation (without shear) is not ruled out by the CMWBR \cite{Obu92},
but this result may be irrelevant to the real universe where shear is
essential to non-trivial perturbations \cite{Goo83,Dun92}; in any case
shearfree models in general relativity are a very restricted class
\cite{Ell67}.

 An example of the problem with assuming a perfect fluid is that in
Bianchi models, as soon as matter is in motion relative to the
homogeneous surfaces (i.e.\ becomes `tilted') it experiences density
gradients which should lead to heat fluxes \cite{BraSvi84}: similar
remarks apply to other simple models. Such models have recently been
used to fit the observed dipole anisotropy in the CMWBR \cite{Tur92},
though other explanations seem to me more credible.

\subsection{Early universe effects}

Galaxy formation in anisotropic models has been studied to see if by
this means one could overcome the well-known difficulties in FLRW
models (without inflation), but with negative results
\cite{PerMatShe72}.

A similar investigation was to see if the helium abundance, as known
in the 1960s, could be fitted better by anisotropic cosmologies than
by FLRW models, which at the time appeared to give discrepancies.  The
reason this might happen is that anisotropy speeds up the evolution
between the time when deuterium can first form, because it is no
longer dissociated by the photons, and the time when neutrons and
protons are sufficiently sparse that they no longer find each other to
combine. Hawking and Tayler (1966)\nocite{HawTay66} were pioneers in
this effort, which continued into the 1980s but suffered some
mutations in its intention.

First the argument was reversed, and the good agreement of FLRW
predictions with data was used to limit the anisotropy during the
nucleogenesis period (see e.g.\ Barrow (1976), Olson
(1977)\nocite{Bar76,Ols77}). Later still these limits were relaxed as
a result of considering the effects of anisotropic neutrino
distribution functions \cite{RotMat82} and other effects on reaction
rates \cite{JusBajGor83}. It has even been shown
\cite{MatVogMad84,Bar84} that strongly anisotropic Bianchi models, not
obeying the limits deduced from perturbed FLRW models, can produce
correct element abundances, though they may violate other constraints
\cite{MatMad85,MatMadVog85}.

\section{Inhomogeneous cosmologies}

\subsection{Self-similar models}

Some of the self-similar models, especially those relevant to modelling
structure formation, are reviewed in much greater detail in a
complementary talk by Carr, so I will give here only a few details of
other cases.

The geometry of the self-similar models first considered in cosmology
is somewhat like that of the
Bianchi models, except that one of the isometries is replaced by a
homothety, that is to say by a vector field satisfying
 $$
 \xi_{(a;b)} = 2kg_{ab}
 $$
 where $k$ is a constant. This class, where the homothety and two
independent symmetries act, was considered by a number of authors
\cite{Ear74,Lum78,Wu81,HanDem84}\footnote{Due to Western
confusion over Chinese name order, Wu Zhong-Chao is sometimes
incorrectly referred to as W.Z. Chao rather than Wu, Z-C.}, and many
details, parallel in nature to those covered by the detailed studies
of Bianchi models, can be found in those works.

More recently Wainwright, Hewitt and colleagues
\cite{HewWaiGoo88,HewWai90,HewWaiGla91}
have considered cases where the homothety has a timelike rather than
spacelike generator. Like the former class, these solutions are in
fact special cases of ``$G_2$ solutions'' (discussed below) with
perfect fluid matter content. It is found that the spatial variations
can be periodic or monotone; the asymptotic behaviour may be a vacuum or
spatially homogeneous model; the periodic cases are unstable to
increases in the anisotropy; and the singularities can be
acceleration-dominated.

\subsection{Spherically symmetric models}

These have a metric
 $$
 ds^2 = -e^{2\nu} dt^2 + e^{2\lambda}dr^2 +R^2(d\theta^2 +
         \sin^2\theta\,d\phi^2)
 $$
 where $\nu$, $\lambda$ and $R$ are functions of $r$ and $t$. The
precise functional forms in the metric depend on the choice of
coordinates and the additional restrictions assumed. It should be
noted that there are so few undetermined functions that a
sufficiently-complicated energy-momentum will fit a totally arbitrary
choice of the remaining functions: in my view this should not be
regarded as a solution, since no equation is actually solved!
\vspace*{15pt}

\parskip=0pt
Some important subcases have been studied, notably:
 \begin{list}{[\arabic{enumi}]}{\setlength{\itemsep}{0pt}
  \setlength{\partopsep}{0pt}  \setlength{\topsep}{0pt}
  \setlength{\parsep}{0pt}\usecounter{enumi}}
 \item{The dust (pressureless perfect fluid) cases, originally studied
by Lema\^{\i}tre, but usually named after Tolman and Bondi;}
 \item{McVittie's 1933 solution representing a black hole in an FLRW universe;}
 \item{The ``Swiss cheese'' model constructed by matching a
Schwarzschild vacuum solution inside some sphere to an exterior FLRW universe;}
 \item{Shearfree fluid solutions \cite{Wym46,KusQvi48,Ste83,McV84};}
 \item{Self-similar solutions, discussed in Carr's contribution at Pont
d'Oye.}
 \end{list}
\parskip=15pt

 Spherically symmetric models, especially Tolman-Bondi, have often
been used to model galactic scale inhomogeneities, in various
contexts.  Galaxy formation has been studied (e.g.\ Tolman (1934),
Carr and Yahil (1990)\nocite{Tol34,CarYah90,Mes91}): Meszaros (1991)
developed a variation on the usual approach by considering the
shell-crossings, with the aim of producing ``Great Wall'' like
structures, rather than the collapse to the centre producing a
spherical cluster or galaxy.  Some authors have used spherically
symmetric lumps to estimate departures from the simple theory of the
magnitude-redshift relations based on a smoothed out
model\footnote{The point is that the beams of light we observe are
focussed only by the matter actually inside the beam, not the matter
that would be there in a completely uniform model.} (e.g.\ Dyer
(1976), Kantowski (1969a) and Newman
(1979)\nocite{Dye76,Kan69a,New79}): note that these works show that
the corrections depend on the choice of modelling, since Newman's
results from a McVittie model differ from the ones based on Swiss
cheese models.  The metrics also give the simplest models of
gravitational lenses\footnote{The very detailed modern work
interpreting real lenses to study various properties of individual
sources and the cosmos mostly uses linearized approximations.} and
have also been used to model the formation of primordial black holes
\cite{CarHaw74}.

On a larger scale, inhomogeneous spherical spacetimes have been used
to model clusters of galaxies \cite{Kan69b}, variations in the Hubble
flow due to the supercluster \cite{Mav77}, the evolution of cosmic
voids \cite{Sat84,HauOlsRot83,BonCha90}\nocite{BonCha91}, the
observed distribution of galaxies and simple hierarchical models of
the universe \cite{Bon72,Wes78,Wes79,Rib92a}. Most of this work used
Tolman-Bondi models, sometimes with discontinuous density
distributions.

Recent work by Ribeiro (1992b)\nocite{Rib92b}, in the course of an
attempt to make simple models of fractal cosmologies using
Tolman-Bondi metrics, has reminded us of the need to compare data with
relativistic models not Newtonian approximations. Taking the
Einstein-de Sitter model, and integrating down the geodesics, he
plotted the number counts against luminosity distances. At small
distances, where a simple interpretation would say the result looks
like a uniform density, the graph is irrelevant because the distances
are inside the region where the QDOT survey shows things are lumpy
\cite{SauFreRow91}, while at greater redshifts the universe ceases to
have a simple power-law relation of density and distance. Thus even
Einstein-de Sitter does not look homogeneous!

One must therefore ask in general ``do homogeneous models look
homogeneous?''. Of course, they will if the data is handled with
appropriate relativistic corrections, but to achieve such comparisons
in general requires the integration of the null geodesic equations in
each cosmological model considered, and, as those who have tried it
know, even when solving the field equations is simple, solving the
geodesic equations may not be.

Many other papers have considered spherically symmetric models, but
there is not enough space here to review them all, so I will end by
mentioning a {\em jeu d'esprit} in
which it was shown that in a ``Swiss cheese'' model, made by joining
two FLRW exteriors at the two sides of a Kruskal diagram for the
Schwarzschild solution, one can have two universes each of which can
receive (but not answer) a signal from the other \cite{Sus85}.

\subsection{Cylindrically symmetric and plane symmetric (static) models}

 These have been used to model cosmic strings and domain walls. One
should note that locally the metrics may be the same for these two
cases, the difference lying in whether there is or is not a Killing
vector whose integral curves have spatial topology $S^1$. Plane
symmetric metrics should have a rotational symmetry in the plane, but
to add to the possible confusions some authors use the term ``plane'' for
solutions without such a rotation: the term ``planar'' would be a
useful alternative.

 The usual (though not the only) form for the cylindrically symmetric
metrics is
 $$
 ds^2 = -T^2dt^2 +R^2dr^2 +Z^2dz^2 +2W\,dz\,d\phi + \Phi^2d\phi^2
 $$
 where $T$, $R$, $Z$, $W$ and $\Phi$ depend on $r$ (and, in the
non-stationary case, $t$) and $\phi$
is periodic, and, for the plane symmetric case,
 $$
 ds^2 = -T^2dt^2 +R^2dr^2 +X^2(dx^2 + x^2d\phi^2)
 $$
 where $T$, $R$ and $X$ are functions of $r$ (and perhaps $t$). The
static cases all belong in Harness's (1982) general class\nocite{Har82}.

Plane symmetric models, usually static, solutions have been used to
model domain walls \cite{Vil83,IpsSik84,Goe90,Wan91a}\footnote{Note
that since the sources usually have a boost symmetry in the timelike
surface giving the wall, corresponding solutions have timelike
surfaces admitting the (2+1)-dimensional de Sitter group.}. The
cylindrically symmetric models have been used for cosmic strings,
starting with the work of Gott\nocite{Got85}, Hiscock\nocite{His85}
and Linet\nocite{Lin85} in 1985. These studies have usually been done
with static strings\footnote{There is some controversy about whether
these can correctly represent strings embedded in an expanding
universe \cite{ClaEllVic90}.}, and have considered such questions as
the effects on classical and quantum fields in the neighbourhood of
the string.

\subsection{$G_2$ cosmologies}

I use the above title as a general name for all cosmological metrics
with two spacelike Killing vectors (and hence two essential
variables). The cylindrical and plane metrics, and many of the Bianchi
metrics, are special cases of $G_2$ cosmologies.
\vspace*{15pt}

\parskip=0pt
$G_2$ cosmologies admit a number of specializations, such as:
 \begin{list}{[\arabic{enumi}]}{\setlength{\itemsep}{0pt}
  \setlength{\partopsep}{0pt}  \setlength{\topsep}{0pt}
  \setlength{\parsep}{0pt}\usecounter{enumi}}
  \item{the Killing vectors commute;}
  \item{the orbits of the $G_2$ are orthogonal to another set of
2-dimensional surfaces $V_2$;}
  \item{the Killing vectors individually are hypersurface-orthogonal;}
  \item{the matter content satisfies conditions allowing generating
techniques.}
 \end{list}
Among the classes of metrics covered here are colliding wave models,
cosmologies with superposed solitonic waves, and what I call
``corrugated'' cosmologies with spatial irregularities dependent on
only one variable.
\parskip=15pt

The metrics where the Killing vectors do {\em not\/} commute have been
very little studied: it is known they cannot admit orthogonal $V_2$ if
the fluid flows orthogonal to the group surfaces (unless they have an
extra symmetry) and that if the fluid is thus orthogonal it is
non-rotating \cite{Bug87,Ber88}. So we now take
only cases where the Killing vectors commute.

The case without orthogonal $V_2$ has also been comparatively little
studied, but recently some exact solutions which have one
hypersurface-orthogonal Killing vector and in which the metric
coefficients are separable, have been derived and studied
\cite{BerWilCas91,Ber91a}. One
class consists of metrics of the form
 $$
 ds^2 = e^{2(K+k)}(-dt^2 + dx^2) +e^{2(S+s)}[(e^{F+f}dy)^2
     +(e^{-(F+f)}\mbox{\boldmath $\theta$})^2]
 $$
 where: $K$, $S$ and $F$ depend on $t$; $k$, $s$ and $f$ depend on $x$;
$\mbox{\boldmath $\theta$} = dz +2\omega\,dx$; and $\omega$ depends on $t$ and
$x$. Some perfect fluid solutions are known explicitly but usually
turned out to be self-similar, with big-bang singularities of the
usual types. The ``stiff fluid'' ($\gamma = 2$) is a special case,
discussed in detail by van den Bergh (1991). Most of the solutions
have singularities at finite spatial distances or can be regarded as
inhomogeneous perturbations of the Bianchi $VI_{-1}$ models.

The cases with orthogonal $V_2$ were classified by Wainwright
(1979;1981)\nocite{Wai79,Wai81}, and a number of specific examples
are known (e.g.\ Wainwright and Goode (1980); Kramer (1984)%
\nocite{WaiGoo80,Kra84}). A recent solution found by Senovilla (1990)
attracted much attention\nocite{Sen90}, because it is non-singular
\cite{ChiFerSen91}, evading the focussing conditions in the
singularity theorems by containing matter that is too diffuse: it is
closely related to an earlier solution of Feinstein and Senovilla
(1989)\nocite{FeiSen89}\footnote{Some recent work has given
generalizations of these solutions \cite{RuiSen92,BerSke92}; also S.W.
Goode at GR13 (unpublished).}. The metrics investigated in this class
generally have Kasner-like behaviour near the singularity (though some
have a plane-wave asymptotic behaviour \cite{Wai83}) and become
self-similar or spatially homogeneous in the far future.

Finally we come to the most-studied class, those where the generating
techniques are applicable. The matter content must have characteristic
propagation speeds equal to the speed of light, so attention is
restricted to vacuum, electromagnetic, neutrino and ``stiff fluid''
(or equivalently, massless scalar field with a timelike gradient)
cases. However, FLRW fluid solutions can be obtained by using the same
methods in higher-dimensions and using dimensional reduction. There
are useful reviews covering the cosmological, cylindrical, and colliding
wave sub-classes \cite{CarChaMal81,Ver85,Ver92,Fer90,Gri91}.
The metrics can be written in a
form covering also the related stationary axisymmetric metrics as
 $$
 ds^2 = \epsilon f_{AB}dx^A dx^B +\delta e^{2\gamma} ((dx^4)^2 -
        \epsilon(dx^3)^2)/f
 $$
 where $A,\,B$ take values $1,\,2$ and
 the values of $f_{AB}$ can be written as a matrix
 $$
 \left( \begin{array}{cc}
 f       &   -f\omega \\
 -f\omega &  f\omega^2+\epsilon(x^3)^2/f
 \end{array} \right)
 $$
 The case $\delta = -\epsilon = 1$ gives the stationary axisymmetric
metrics, the case $\delta = \epsilon = 1$ the cylindrical cases and
$\epsilon = -\delta =1$ the cosmological cases.
Physically these classes differ in the timelike or
spacelike nature of the surfaces of symmetry and the nature of the
gradient of the determinant of the metric in those surfaces.

Some studies have focussed on the mathematics, showing how known
vacuum solutions can be related by solution-generating techniques
\cite{Kit84}, while others have concentrated on the physics of the
evolution and interpretative issues.  The generating techniques use
one or more of a battery of related methods: B\"acklund
transformation, inverse scattering, soliton solutions and so on. One
interesting question that has arisen from recent work is whether
solitons in relativity do or do not exhibit non-linear interactions:
Boyd {\em et al.\/} (1991)\nocite{BoyCenKla91}, in investigations of
solitons in a Bianchi I background, found no non-linearity, while
Belinskii (1991)\nocite{Bel91} has claimed there is a non-linear
effect (see also Verdaguer (1992)).

The applications in cosmology, which have generated far too many papers
to list them all here, have been pursued by a number of groups,
notably by Carmeli,
Charach and Feinstein, by Verdaguer and colleagues, by Gleiser, Pullin
and colleagues, and Belinski, Curir and Francaviglia, with important
contributions by Ibanez, Kitchingham, Yurtsever, Ferrari,
Chandrasekhar and Xanthopoulos, Letelier, Tsoubelis and Wang and many
others.

One use of these metrics is to provide models for universes with
gravitational waves.  It emerges that the models studied are typically
Kasner-like near the singularity (agreeing with the LK arguments), and
settle down to self-similar or spatially homogeneous models with
superposed high-frequency gravitational waves at late times
\cite{AdaHelZim82,CarFei84,Fei88}. Another use
is to model straight cosmic strings in interaction with gravitational
or other waves (e.g.\ Economou and Tsoubelis (1988)\nocite{EcoTso88},
Verdaguer (1992)).
One can also examine the gravitational analogue of Faraday rotation
\cite{PirSaf85,Tom89,Wan91b} and there are even solutions
whose exact behaviour agrees precisely with the linearized
perturbation calculations for FLRW universes \cite{CarChaFei83}.

\subsection{Other models}

Solutions with less symmetry than those above have been little
explored. Following Krasinski one can
divide the cases considered into a number of classes (in which I only
mention a few important special subcases).

\begin{enumerate}
\item{The Szekeres-Szafron family (also independently found by
Tomimura). These have in general no symmetries, and contain an
irrotational non-accelerating fluid. Tolman-Bondi universes
are included in this class, as are the Kantowski-Sachs metrics;
some generalizations are known, such as the
rotating inhomogeneous model due to Stephani (1987)\nocite{Ste87}.
Like the $G_2$ solutions mentioned earlier, some Szekeres models obey
exactly the linearized perturbation equations for the FLRW models
\cite{GooWai82}.}

\item{Shearfree irrotational metrics \cite{Bar73} which include
the conformally flat fluids (Stephani 1967a,
1967b)\nocite{Ste67a,Ste67b} and McVittie's spherically symmetric
metric.  Bona and Coll (1988)\nocite{BonCol88} have recently argued
that the Stephani cases can only have acceptable thermodynamics if the
metrics admit three Killing vectors. }

 \item{The Vaidya-Patel-Koppar family, which represent an FLRW model
contaning a ``Kerr'' solution using null radiation and an
electromagnetic field. The physical significance of these metrics is dubious.}

 \item{Some other special cases such as Oleson's Petrov type N fluid
solutions.}
 \end{enumerate}

\section{Syntheses and conclusions: what have we learnt?}

Here I collect up the outcome of the work surveyed above, without
repeating all details, and review some relevant extra references, but
many interesting aspects are still omitted. For example, the
literature covers such issues as models for interactions between
different forms of matter, and generation of gravitational radiation.

\subsection{The classical singularity}

 The occurrence of a ``big-bang'' in FLRW models is not just a
consequence of the high symmetry. Its nature in general models is
probably a curvature singularity, and the best guess so far is that
the asymptotic behaviour would be oscillatory but other
possibilities exist. The Penrose conjecture, which would be a
selection principle on models, has been particularly developed, using
exact solutions as examples, by Wainwright and Goode, who have given a
precise definition of the notion of an `isotropic singularity'
\cite{GooColWai92,Tod92}.

\subsection{Occurrence of inflation}

In ``old'' inflation in Bianchi I models, inflation need not occur
\cite{BarTur81}, but in ``new'' inflation it was predicted
\cite{SteTur83}. In a large class of chaotic inflation
models it is also expected \cite{MosSah86}. Further papers by a
number of authors have suggested that inflation need not always occur
(see Rothman and Ellis (1986)\nocite{RotEll86} for some criticisms of
earlier papers).

\subsection{Removal of anisotropy and inhomogeneity}

Three means of smoothing the universe have been explored over the
years: the use of viscosity in the early universe; the removal of
horizons in the Mixmaster universes; and removal during inflation. The
first two of these ingenious suggestions are due to Misner.

Attempts to smooth out anisotropies or inhomogeneities by any process
obeying deterministic sets of differential equations satisfying
Lipschitz-type conditions are doomed to fail, as was first pointed out
by Collins and Stewart (1971)\nocite{ColSte71} in the context of
viscous mechanisms. The argument is simply that one can impose any
desired amount of anisotropy or inhomogeneity now and evolve the
system backwards in time to reach initial conditions at some earlier
time whose evolution produces the chosen present-day values.

The same argument also holds for inflationary models. Inflation in
itself, without the use of singular equations or otherwise
indeterminate evolutions, cannot wholly explain present isotropy or
homogeneity, although it may reduce deviations by large factors
\cite{Sir82,Wal83,MosSah86,FutRotMat89}. Although one can argue that
anisotropy tends to prolong inflation, this does not remove the
difficulty.

Since 1981 I have been arguing a heretical view about one of the
grounds for inflation, namely the `flatness problem', on the grounds
that the formulation of this problem makes an implicit and unjustified
assumption that the {\em a priori} probabilities of values of $\Omega$ is
spread over some range sufficient to make the observed closeness to 1
implausible. Unless one can justify the {\em a priori} distribution, there
is no implausibility\footnote{One can however argue that only
$\Omega\:=\:1$ is plausible, on the grounds that otherwise the quantum
theory before the Planck time would have to fix a length-scale
parameter much larger than any quantum scale, only the $\Omega\:=\:1$
case being scale-free. I am indebted to Gary Gibbons
for this remark.} \cite{Ell91}.

However, if one accepts there is a flatness problem, then there is
also an isotropy problem, since at least for some probability
distributions on the inhomogeneity and anisotropy the models would
not match observation. Protagonists of inflation cannot have it both
ways. Perhaps, if one does not want to just say ``well, that's how the
universe was born'', one has to explain the observed smoothness by
appeal to the `speculative era', as Salam (1990) called it\nocite{Sal90},
i.e.\ by appeal to one's favourite theory of quantum gravity.

 If inflation works well at early times,
then inflation actually enhances the chance of an anisotropic model fitting the
data, and since the property of anisotropy cannot be totally
destroyed in general (because it can be coded into geometric invariants
which cannot become zero by any classical evolution) the anisotropy
could reassert itself in the future! (This of course will not happen
if a non-zero $\Lambda$ term persists, as the ``cosmic no-hair''
theorems show \cite{Wal83,MorWit88}.)

The Mixmaster horizon removal suggestion was shown to fail
when more detailed computations than Misner's were made
\cite{Mac71,DorLukNov71,Chi72}.
Incidentally, one may note that inflation does not solve the original
form of the `horizon problem', which was to account completely for the
similarity of points on the last scattering surface governed by
different subsets of the inital data surface. Inflation leads to a
large overlap between these initial data subsets, but not to their
exact coincidence. Thus one still has to assume that the non-overlap
regions are not too different. While this may give a more plausible
model, it does not remove the need for assumptions on the initial
data.

\subsection{The exit from inflation}

A further interesting application of non-standard models has come in a
recent attempt to answer the question posed by Ellis and Rothman
(unpublished) of how the universe can choose a uniform reference frame
at the exit from inflation when a truly de Sitter model has no
preferred time axis. Anninos {\em et al.\/}
(1991a)\nocite{AnnMatRot91} have shown by taking an inflating Bianchi V
model that the answer is that the memory is retained and the universe
is never really de Sitter.

\subsection{The helium abundance}

This is still used to set limits on anisotropy during the
nucleosynthesis phase.

\subsection{The cosmic microwave background}

Observations limit the integrated effect since ``last scattering'':
note this can in principle permit large but compensating excursions
from FLRW. One intriguing possibility raised by Ellis {\em et al.}
(1978)\nocite{EllMaaNel78} is that the observed sphere on the last
scattering surface could lie on a timelike (hyper)cylinder of
homogeneity in a static spherically symmetric model. This makes the
CMWBR isotropic {\em at all points} not only at the centre, and
although it cannot fit all the other data, the model shows how careful
one must be, in drawing conclusions about the geometry of the universe
from observations, not to assume the result one wishes to prove.

There is a theorem by Ehlers, Geren and Sachs
(1968)\nocite{EhlGerSac68} showing that if a congruence of
geodesically-moving observers all observe an isotropic distribution of
collisionless gas the metric must be Robertson-Walker. Treciokas and
Ellis (1971) have investigated the related problem with
collisions\nocite{TreEll71}. Recently Ferrando {\em et al.\/} (1992)
have investigated inhomogeneous models where an isotropic gas
distribution is possible\nocite{FerMorPor92}. These studies throw
into focus a conjecture which is usually assumed, namely that an
approximately isotropic gas distribution, at all points, would imply
an approximately Robertson-Walker metric. (It is this assumption which
underlies the arguments normally used in analysing data like that from
COBE to get detailed information on allowed FLRW perturbations.)

\subsection{The far future}

Anisotropy will in general become apparent, if it is present and if the
cosmological constant $\Lambda$ is zero: isotropy is not stable.
Inhomogeneities may become significant even faster.

\subsection{The origin of structure}

None of the work discussed above accounts for the origin of structure,
although it offers suitable descriptions for the evolution, or the
background spacetime in which the evolution takes place. I feel it
does, however, indicate strongly that the true origin lies in the
perhaps unknowable situation in the Speculative Era, and the resulting
initial conditions for the later evolution.

\subsection{A genuinely anisotropic and inhomogeneous universe?}

While I do not think one can give a definitive answer to this
question, I would personally be very surprised if anisotropic but
homogeneous models turned out to be anything more than useful
examples. However, the status of fully inhomogeneous models is less
clear.

One argument is that while the standard models may be good
approximations at present, they are unstable to perturbations both in
the past and the future. The possible alternative pasts are quite
varied, as shown above, even without considering quantum
gravity. Similarly, the universe may
not be isotropic in the far future. Moreover, we have no knowledge of
conditions outside our past null cone, where some inflationary
scenarios would predict bubbles of differing FLRW universes, and
perhaps domain walls and so on.

If the universe were FLRW, or very close to that, this means it is in a
region, in the space of all possible models, which almost any
reasonable measure is likely to say has very low probability (though
note the earlier remarks on assignments of probabilities). One
can only evaluate, and perhaps explain, this feature by considering
non-FLRW models. It is noteworthy that many of the ``problems''
inflation claims to tackle are not problems if the universe simply is
always FLRW. Hence, as already argued above, one has a deep problem in
explaining why the universe is in the unlikely FLRW state if one
accepts the arguments about probabilities current in work on inflation.

Suppose we speculated that the real universe is significantly
inhomogeneous at the present epoch (at a level beyond that arising
from perturbations in FLRW). What would the objections be? There are
only two relevant pieces of data, as far as I can see. One is the deep
galaxy counts made by the automatic plate measuring machines, which
are claimed to restrict variations to a few percent, and the other is
the isotropy of the CMWBR. Although the latter is a good test for
large lumps in a basically FLRW universe, one has to question
(recalling the results of Ellis {\em et al.\/} (1978)\nocite{EllMaaNel78})
whether it really implies homogeneity.



\end{document}